\documentclass[12pt,thmsa]{article}
\usepackage{sw20lart}

\input{tcilatex}
\begin{document}

\author{Vivian de la Incera \\
%EndAName
Physics Dept., SUNY-Fredonia, Fredonia, NY 14063}
\title{External Magnetic Fields in QFT: A Non-Perturbative Approach}
\date{{\small Talk given at ''Quantization, Gauge Theory and Strings, dedicated to
the memory of E. S. Fradkin'', Moscow, June 5-10, 2000}}
\maketitle

\begin{abstract}
A discussion of the influence of boundaries and scalar field interactions in
the non-perturbative dynamics of fermions in an external magnetic field,
along with their possible applications to condensed matter and cosmology, is
briefly presented. The significance of the results for electroweak
baryogenesis in the presence of an external magnetic field is indicated.
\end{abstract}

\section{Introduction}

In this talk I would like to present some recently found non-perturbative
effects of external magnetic fields in quantum field theories with fermions.
They are related to the so called Magnetic Catalysis (MC) \cite
{mirans-gus-shoko}, a universal phenomenon that can be understood as the
generation, through the infrared dynamics of the fermion pairing in a
magnetic field, of a fermion dynamical mass (fermion gap) at the weakest
attractive interaction between fermions.

The discussion will be particularized to the following two problems: (1) the
influence of boundaries in the dynamical generation of a fermion mass in the
presence of a magnetic field and (2) the modification of the nonperturbative
fermion dynamics due to fermion-scalar interactions, and its possible
implications for baryogenesis in the presence of strong magnetic fields.

\section{Boundary Effects}

The influence of non-trivial boundaries on the MC was investigated by
Ferrer, Gusynin and Incera \cite{FGI-bound}, who considered the Nambu-Jona
Lasinio (NJL) model in a locally flat space-time with topology represented
by the domain $R^{3}\times S^{1}$ and in the presence of an external
constant magnetic field.

Starting from the NJL effective action

\begin{equation}
W(\sigma ,\pi )=-\frac{N}{2G}\int d^{4}x(\sigma ^{2}+\pi ^{2})-i\mathrm{Tr}%
\log \left[ i\gamma ^{\mu }D_{\mu }-\left( \sigma +i\gamma _{5}\pi \right)
\right] ,  \label{1}
\end{equation}
where the composite scalar $\sigma $ and $\pi $ are defined by

\begin{equation}
\sigma =-\frac{G}{N}\left( \overline{\psi }\psi \right) ,\qquad \pi =-\frac{G%
}{N}\left( \overline{\psi }i\gamma ^{5}\psi \right) ,  \label{2}
\end{equation}
one can investigate the generation of a fermion dynamical mass by
determining the stationary points of $W.$ Since the vacuum should respect
translational invariance, it is enough to calculate the effective action for
constant auxiliary fields. In this case, the effective action is just $%
W(\sigma ,\pi )=-V(\sigma ,\pi )T\mathcal{A}a$, where $T\mathcal{A}a$ is the
space-time volume and $V$ is the effective potential. In the proper-time
formalism one gets

\begin{eqnarray}
V_{AP}(\sigma ) &=&\frac{N\sigma ^{2}}{2G}+\frac{NeB}{8\pi ^{2}}%
\int\limits_{1/\Lambda ^{2}}^{\infty }\frac{ds}{s^{2}}e^{-s\sigma
^{2}}\theta _{4}\left( 0\left| \frac{ia^{2}}{4\pi s}\right. \right) \coth
(eBs),  \label{VAP} \\
V_{P}(\sigma ) &=&\frac{N\sigma ^{2}}{2G}+\frac{NeB}{8\pi ^{2}}%
\int\limits_{1/\Lambda ^{2}}^{\infty }\frac{ds}{s^{2}}e^{-s\sigma
^{2}}\theta _{3}\left( 0\left| \frac{ia^{2}}{4\pi s}\right. \right) \coth
(eBs),  \label{VP}
\end{eqnarray}
for antiperiodic \textit{(APBC)} and periodic \textit{(PBC)}\ boundary
conditions of the fermion fields at the boundary respectively\cite{FGI-bound}%
.

The dynamical mass $\bar{\sigma}$ for the \textit{PBC} case is the solution
of the gap equation $dV_{P}/{d\sigma }=0$. In leading order in $1/\Lambda $
the PBC gap equation is given by 
\begin{eqnarray}
&&\sigma \left[ {\frac{1}{G}}-\frac{\Lambda ^{2}}{4\pi ^{2}}+\frac{\sigma
^{2}}{4\pi ^{2}}\left( \log \frac{\Lambda ^{2}}{\sigma ^{2}}+1-\gamma
\right) -\frac{1}{4\pi ^{2}}\int\limits_{0}^{\infty }\frac{ds}{s^{2}}%
e^{-s\sigma ^{2}}\left( \theta _{3}\left( 0\left| \frac{ia^{2}}{4\pi s}%
\right. \right) -1\right) \right.  \nonumber \\
&&\left. -\frac{eB}{4\pi ^{2}}\int\limits_{0}^{\infty }\frac{ds}{s}%
e^{-s\sigma ^{2}}\theta _{3}\left( 0\left| \frac{ia^{2}}{4\pi s}\right.
\right) \left( \coth (eBs)-{\frac{1}{eBs}}\right) +O\left( \frac{1}{\Lambda }%
\right) \right] =0,  \label{gapequation}
\end{eqnarray}
where $\gamma \approx 0.577$ is the Euler constant. The corresponding gap
equation for \textit{APBC} is obtained by replacing $\theta _{3}$ by $\theta
_{4}$ in Eq. (\ref{gapequation}).

It is easy to see that under the condition $1/a\ll \sigma \ll \sqrt{eB}$,
i.e. $a$ being the largest length scale in the problem, the gap equations
for both cases (\textit{PBC }and \textit{APBC}) reduce to the following one 
\begin{equation}
\sigma \left[ \frac{1}{G}-\frac{1}{G_{c}}\pm \left( \frac{2\sigma }{\pi a^{3}%
}\right) ^{1/2}e^{-\sigma a}-\frac{eB}{4\pi ^{2}}\log \frac{eB}{\pi \sigma
^{2}}\right] =0,  \label{AA}
\end{equation}
where $G_{c}=({4\pi ^{2}}/\Lambda ^{2})$ and $\pm $ refers to \textit{APBC}
and \textit{PBC}, respectively. The solution of Eq. (\ref{AA}) in the $G\ll
G_{c}$ approximation is 
\begin{equation}
m_{dyn}\equiv \bar{\sigma}=\sqrt{\frac{eB}{\pi }}\exp \left( -\frac{2\pi ^{2}%
}{eBG}\right) .  \label{PLmass}
\end{equation}
As expected, the solution (\ref{PLmass}), which is nonanalytic in $G$ as $%
G\to 0,$ coincides with the one found in (3+1)-dimensions for $B\neq 0$ and $%
a=\infty $ \cite{mirans-gus-shoko}.

Let us consider now the small length $a$ ($\sigma ,\sqrt{eB}\ll 1/a$) limit,
which is the important one to study the effects of the compactified
dimension. In this case the gap equation (\ref{gapequation}) for \textit{PBC}
reduces to 
\begin{equation}
\sigma \left[ {\frac{1}{G}}-\frac{1}{G_{c}}-\frac{1}{3a^{2}}+\frac{\sigma
^{2}}{4\pi ^{2}}\left( \log \frac{\Lambda ^{2}a^{2}}{16\pi ^{2}}+\gamma
\right) -\frac{eB}{2\pi \sigma a}\left( \sqrt{\frac{2}{eB}}\sigma \zeta ({%
\frac{1}{2}},\frac{\sigma ^{2}}{2eB}+1)+1\right) \right] =0,
\label{gapeq:PBC}
\end{equation}
where $\zeta (\nu ,x)$ is the generalized Riemann zeta function.

As $B\rightarrow 0$, ones recovers the known gap equation \cite{Klimenko}
which admits a nontrivial solution only if the coupling is supercritical, $%
G>G_{c}^{a}$, and the critical coupling $%
G_{c}^{a}=(G_{c}^{-1}+1/3a^{2})^{-1} $. When an external magnetic field, $%
B\neq 0$, is present a nontrivial solution exists at all $G>0$ and, in
particular, at $G\ll G_{c}^{a}$. Indeed, looking at the solution of Eq. (\ref
{gapeq:PBC}) satisfying $\bar{\sigma}\ll \sqrt{eB}$, it is found 
\begin{equation}
m_{dyn}^{\mathit{P}}\equiv \bar{\sigma}\simeq \frac{eB}{2\pi a}\frac{%
GG_{c}^{a}}{G_{c}^{a}-G}  \label{3}
\end{equation}
if the coupling $G\ll G_{c}^{a}$. The condition $G<G_{c}^{a}$ guarantees
that (\ref{3}) is a minimum solution of the effective potential $V_{P}$.

Therefore a dynamical mass solution (\ref{3}) exists in the weak coupling
regime of the theory. The fact that there is no critical value of the
coupling to produce chiral symmetry breaking is a characteristic feature of
the catalysis of dynamical symmetry breaking by a magnetic field\cite
{mirans-gus-shoko}. It is remarkable that unlike the $a=\infty $ case, where
the dynamical mass has nonanalytical dependence on the coupling constant as $%
G\to 0$, at finite $a$ the dynamical mass (\ref{3}) is an analytic function
of $G$ at $G=0$.

Note also that $m_{\mathrm{dyn}}=\langle 0|\sigma |0\rangle =-G\langle 0|%
\bar{\psi}\psi |0\rangle /N$. From here and Eq. (\ref{3}) one finds that the
condensate $\langle 0|\bar{\psi}\psi |0\rangle $ is $\langle 0|\bar{\psi}%
\psi |0\rangle =-N|eB|/{2\pi a}$ in leading order in $G$; i.e. it coincides
with the value of the condensate calculated in the problem of free fermions
in a magnetic field (see \cite{FGI-bound}). This point also explains why the
dynamical mass $m_{\mathrm{dyn}}$ is an analytic function of $G$ at $G=0$:
indeed, the condensate already exists at $G=0$. As a result, there is a big
enhancement of the dynamical fermion mass generation in the $R^{3}\times
S^{1}$ domain with periodic boundary conditions for the fermion fields as
compared to the case of topologically trivial space-time (see Eqs. (\ref{3})
and (\ref{PLmass})).

Let us discuss now the \textit{APBC} case. Following the same procedure used
for \textit{PBC}, it is easy to show that the \textit{APBC} gap equation
under conditions $\sigma ,\sqrt{eB}\ll 1/a$ does not have a nontrivial
solution; on the other hand, when $\sigma \ll \sqrt{eB},1/a$ it is reduced
to 
\begin{eqnarray}
&&\sigma \left[ {\frac{1}{G}}-\frac{1}{G_{c}}+\frac{1}{6a^{2}}-\frac{eB}{%
4\pi ^{2}}\int\limits_{0}^{\infty }\frac{ds}{s}\theta _{4}\left( 0\left| 
\frac{i}{4\pi s}\right. \right) \left( \coth (eBa^{2}s)-{\frac{1}{eBa^{2}s}}%
\right) \right.   \nonumber \\
&&+\left. \frac{\sigma ^{2}}{4\pi ^{2}}\left( \log \frac{\Lambda ^{2}a^{2}}{%
\pi ^{2}}+\gamma +eBa^{2}\int\limits_{0}^{\infty }ds\theta _{4}\left(
0\left| \frac{i}{4\pi s}\right. \right) \left( \coth (eBa^{2}s)-{\frac{1}{%
eBa^{2}s}}\right) \right) \right] =0.  \nonumber \\
&&  \label{4}
\end{eqnarray}
From Eq. (\ref{4}) one can convince oneself that there is no nontrivial
solution under the assumptions made if the coupling is weak ($G\to 0$). For
chiral symmetry breaking to take place, the coupling constant $G$ must be
larger than some critical value that depends on the magnitude of the
magnetic field $B$ and size $a$. Indeed, Eq.(\ref{4}) can be simplified in
the limiting case $\sqrt{eB}a\gg 1$ 
\begin{equation}
\sigma \left[ \frac{1}{G}-\frac{1}{G_{c}}+\frac{1}{6a^{2}}-\frac{eB}{4\pi
^{2}}\left( \ln \frac{eBa^{2}}{\pi ^{3}}+2\gamma \right) +\frac{\sigma ^{2}}{%
4\pi ^{2}}\left( \log \frac{\Lambda ^{2}a^{2}}{\pi ^{2}}+\frac{7\zeta (3)}{%
4\pi ^{2}}eBa^{2}+\gamma \right) \right] =0.  \label{5}
\end{equation}
From Eq. (\ref{5}) one can notice that the magnetic field is helping the
symmetry breaking since the critical coupling is less than the one
corresponding to the case with zero magnetic field.

On the other hand, the contribution of the compactified dimension length $a$
to the gap equation in the presence of a magnetic field has opposite sign
for \textit{APBC} (third term in Eq. (\ref{4})), as compared to \textit{PBC}
(third term in Eq. (\ref{gapeq:PBC})). Consequently, the boundary effect in
the \textit{APBC} case is not enhancing the chiral symmetry breaking, but on
the contrary, it is counteracting it; while in the \textit{PBC} case the
magnetic catalysis is substantially enhanced by the boundary.

The behavior of the system in the \textit{APBC} case can be better
understood by realizing that due to the antiperiodic boundary conditions the
quantity $1/a$ plays a role similar to temperature. Notice that chiral
symmetry breaking takes place at small $1/a$ with a corresponding dynamical
mass given by Eq. (\ref{PLmass}), so one should expect that at $1/a$ larger
than some critical value $1/a_{c}$ the chiral symmetry must be restored.

Such a critical value $1/a_{c}$ indeed exists and is determined from the
condition that the second derivative of the effective potential at $\sigma
=0 $ becomes positive. In fact, from Eq. (\ref{5}) it is found to be 
\begin{equation}
\frac{1}{a_{c}}=\frac{e^{\gamma }}{\pi }m_{dyn}  \label{ac}
\end{equation}
with $m_{dyn}$ the dynamical mass (\ref{PLmass}). Therefore, one obtains
that the inverse of the critical length, $1/a_{c}$, is of the order of $%
m_{dyn}$, a result equivalent to that found for the relationship between the
critical temperature and the gap in BCS superconductivity.

Finally, one should point out that the combined effect of an external
magnetic field and non-trivial boundary conditions for fermions along third
axis can find applications in condensed matter, in particular in high-$T_{c}$
superconductors which are known to possess a quasi-2D structure (see
discussion in \cite{FGI-bound}).

\section{Scalar Interactions}

Let us now investigate the influence of scalar fields in the
non-perturbative dynamics of fermions in a magnetic field background \cite
{MC-sca}. With this goal in mind, let us consider a model field theory
containing self interacting scalar fields, as well as Yukawa fermion-scalar
interactions in the presence of an external constant magnetic field. As seen
below, due to the magnetic field, a non-zero vacuum expectation value (vev)
of the scalar field and a fermion dynamical mass arise as solution of the
minimum equations of the system, breaking in this way the discrete chiral
symmetry of the original Lagrangian. However, in contrast to more
conventional mechanisms to generate scalar vev's, the present model does not
require the introduction of a scalar mass term with a wrong sign in the
original Lagrangian, nor does it need dimensional transmutation ''a la
Coleman-Weinberg.'' Instead, the scalar vev is catalyzed, along with a
fermion-antifermion condensate, by the external magnetic field. Both the
fermion condensate and the scalar vev contribute to the dynamically
generated fermion mass. The scalar vev grows with the square root of the
magnetic field strength. No particular value of the magnetic field needs to
be assumed, since there is no critical magnetic field for the
zero-temperature symmetry breaking considered in the present work. The only
assumption made is that all couplings are weak enough to justify a
Hartree-Fock approach.

The starting point is a theory of gauge, fermion and scalar fields described
by the following Lagrangian density

\begin{equation}
L=-\frac{1}{4}F^{\mu \nu }F_{\mu \nu }+i\overline{\psi }\gamma ^{\mu
}\partial _{\mu }\psi +g\overline{\psi }\gamma ^{\mu }\psi A_{\mu }-\frac{1}{%
2}\partial _{\mu }\varphi \partial ^{\mu }\varphi -\frac{\lambda }{4!}%
\varphi ^{4}-\frac{\mu ^{2}}{2}\varphi ^{2}-\lambda _{y}\varphi \overline{%
\psi }\psi  \label{e1}
\end{equation}
It has a U(1) gauge symmetry, a fermion number global symmetry, and the
discrete chiral symmetry 
\begin{equation}
\psi \rightarrow \gamma _{_{5}}\psi ,\qquad \overline{\psi }\rightarrow -%
\overline{\psi }\gamma _{_{5}},\qquad \varphi \rightarrow -\varphi
\label{e2}
\end{equation}

To study the vacuum solutions of the theory (\ref{e1}) under the influence
of an external constant magnetic field $B$, we need to solve the extremum
equations of the effective action $\Gamma $ for composite operators\cite
{jackiw}

\begin{eqnarray}
\frac{\delta \Gamma (\varphi _{c},\overline{G})}{\delta \overline{G}} &=&0,\;
\label{e3} \\
\frac{\delta \Gamma (\varphi _{c},\overline{G})}{\delta \varphi _{c}} &=&0
\label{e4}
\end{eqnarray}
where $\overline{G}(x,x)=\sigma (x)=\left\langle 0\mid \overline{\psi }%
(x)\psi (x)\mid 0\right\rangle $ is a composite fermion-antifermion field,
and $\varphi _{c}$ represents the vev of the scalar field. If the minimum
solutions of (\ref{e3}) and (\ref{e4}) are different from zero the discrete
chiral symmetry (\ref{e2}) will be broken and the fermions will acquire
mass. We shall see that due to the magnetic field this is indeed the case.

The leading contributions to Eqs. $\left( \ref{e3}\right) $ and $\left( \ref
{e4}\right) $ at large magnetic field, lead to the following minimum
equations for the fermion mass $m$ and the scalar vev respectively\cite
{MC-sca}, 
\begin{equation}
m\simeq m_{0}+\left( \frac{g^{2}}{4\pi }-\frac{\lambda _{y}^{2}}{8\pi }%
\right) \frac{m}{4\pi }\ln ^{2}\left( \frac{gB}{m^{2}}\right) +\frac{1}{\pi
^{2}}\frac{\lambda _{y}^{2}}{\lambda \varphi _{c}^{2}}gBm\ln \left( \frac{gB%
}{m^{2}}\right)  \label{e5}
\end{equation}
\begin{equation}
\frac{\lambda }{6}\varphi _{c}^{3}+\frac{\lambda ^{2}}{64\pi ^{2}}\varphi
_{c}^{3}\left( \ln \left( \frac{\varphi _{c}^{2}}{gB}\right) -\frac{11}{3}%
\right) -\lambda _{y}\frac{gB}{2\pi ^{2}}m\ln \left( \frac{gB}{m^{2}}\right)
\simeq 0  \label{e6}
\end{equation}

Assuming $\varphi _{c}\ll \sqrt{gB}$ (something that is corroborated by the
results), the solution of these equation can be reduced to 
\begin{equation}
m\simeq \frac{1}{\sqrt{\kappa }}\sqrt{gB}  \label{e7}
\end{equation}
\begin{equation}
\varphi _{c}\approx \frac{0.8}{\kappa ^{1/2}\lambda _{y}}\sqrt{gB}
\label{e8}
\end{equation}
where the coefficient $\kappa $ satisfies

\begin{equation}
\kappa \ln \kappa \simeq 1.4\frac{\lambda }{\lambda _{y}^{4}}  \label{e9}
\end{equation}

Notice that the solutions $\left( \ref{e7}\right) $ and $\left( \ref{e8}%
\right) $ are indeed non-perturbative in the couplings constants. At each
fixed scalar self-coupling $\lambda ,$ the values of $m$ and $\varphi _{c}$
increase with $\lambda _{y},$ since the parameter $\kappa $ falls down much
more rapid than $\frac{1}{\lambda _{y}}$ as $\lambda _{y}$ increases.

It is a well known fact that in the absence of a magnetic field, the
one-loop effective action (effective potential) of the present model would
have a minimum at some non-trivial value of the scalar vev, but this minimum
would lie far outside the expected range of validity of the one-loop
approximation, even for arbitrarily small coupling constant, so it would
have to be rejected as an artifact of the used approximation. In the present
case however, thanks to the magnetic field, a consistent scalar field
minimum solution is generated by non-perturbative radiative corrections. In
this sense, a sort of non-perturbative Coleman-Weinberg mechanism takes
place, with the difference that here no dimensional transmutation occurs.
Since the theory already contains a dimensional parameter: the magnetic
field $B$, there is no need to include scalar-gauge interactions in order to
trade a dimensionless coupling for the dimensional parameter $\varphi _{c}$.

From the obtained results it can be seen that the scalar field interactions
significantly enhance the magnetic catalysis. The enhancement of the
dynamical mass due to the scalar field interactions is comparable to the
effect produced by lowering the number of spatial dimensions in a theory
that already exhibits magnetic catalysis in 3+1 dimensions \cite{NJL-2dimen}.

A noteworthy feature of the results is that there is no trivial solution
(stable or unstable) for the scalar vev in the present theory. Besides, no
critical value of the magnetic field is required to produce the fermionic
condensate and the scalar vev.

When studying this model at finite temperature, it is logical to expect that
the increase in the dynamical mass due to the scalar field interactions will
lead to a corresponding increase in the temperature at which the discrete
symmetry is restored, since, typically, the critical temperature separating
chiral broken-unbroken phases is of the order of the zero-temperature
dynamical mass.

The substitution of real scalars by complex scalars does not lead to any
qualitative change in the main results here discussed. However, the model
with complex scalars better resembles a simplified version of the
electromagnetic sector of the $SU(2)\times U(1)$ electroweak model (the
scalar is not coupled to the electromagnetic field, just as the Higgs field
of the electroweak theory). Since no tree-level scalar mass term with the
wrong sign is introduced, the external magnetic field catalyzes gauge
symmetry breaking by producing a non-trivial scalar vev through the
non-perturbative mechanism shown in this paper. Even more, it is natural to
expect that in the usual situation, where the gauge symmetry is broken in
the standard way through the introduction of a wrong-sign scalar mass in the
Lagrangian, the non-perturbative contributions induced by the external
magnetic field will still yield an increment of the scalar field vev whose
physical consequences are yet to be determined.

How significant for cosmology the non-perturbative effects here discussed
are is still an open question. One can envision though that in the high
temperature region of the electroweak theory, where the
temperature-dependent fermion masses are nearly zero, these magnetic
field-driven non-perturbative effects may be important at reasonable large
magnetic fields. If that is the case, the scalar vev may be consequently
visibly augmented (an effect up to now ignored by previous works). An
interesting point would be then to determine whether the new magnetic field
dependence of the scalar vev is large enough to change the recent
conclusions on the role of magnetic fields in electroweak baryogenesis \cite
{Shap}\cite{Skalaz}.

$\mathbf{Acknowledgments}$

This research has been supported in part by the National Science Foundation
under Grants No. PHY-9722059 and PHY-9973708.

\end{document}